# Action mechanism of DDX3X:

# An RNA helicase implicated in

# cancer propagation and viral infection

Anthony F. T. Moore, Aliana López de Victoria and Eda Koculi[a]

University of Central Florida, Department of Chemistry, 4111 Libra Dr., Physical Sciences Bld. Rm. 255, Orlando, FL 32816-2366

[a.] Corresponding author. E-mail: eda.koculi@ucf.edu; Phone: (407)823-5451.

**ABSTRACT**

DDX3X is a human DEAD-box RNA helicase implicated in many cancers and in viral progression. In addition to the RecA-like catalytic core, DDX3X contains N- and C-terminal domains. Here, we investigate the substrate and protein requirements to support the ATPase activity of a DDX3X construct lacking 80 residues from its C-terminal domain. Our data confirmed previous results that for an RNA molecule to support the ATPase activity of DDX3X it must contain a single-stranded–double-stranded region. We investigated protein and RNA structural reasons for this requirement. First, the RNA substrates consisting only of a double-helix were unable to support DDX3X binding. A single-stranded RNA substrate supported DDX3X binding, while an RNA substrate consisting of a single-stranded–double-stranded region not only supported the binding of DDX3X to RNA, but also promoted DDX3X trimer formation. Thus, the single-stranded–double-stranded RNA region is needed for DDX3X trimer formation, and trimer formation is required for ATPase activity. Interestingly, the dependence of ATP hydrolysis on the protein concentration suggests that the DDX3X trimer hydrolyzes only two molecules of ATP. Lastly, a DNA substrate that contains single-stranded–double-stranded regions does not support the ATPase activity of DDX3X.

**KEYWORDS**

DEAD-box; RNA helicase; ATPase; HIV; Cancer

**INTRODUCTION**

DDX3X belongs to the DEAD-box RNA helicase family of enzymes. DEAD-box RNA helicases, so named because of their conserved amino acid sequence, use the energy of ATP binding and hydrolysis to unwind short RNA double-helixes.[1,2] DDX3X is involved in RNA transcription, transport, translation initiation and ribosome assembly. Consequently, DDX3X is implicated in breast, cervical, prostate, brain, lung, liver, ovarian, prostate cancers and Ewing sarcoma.[3-6] Moreover, DDX3X is involved in HIV-1 and Hepatitis C propagation.[7-12] Despite DDX3X involvement in crucial cellular functions and its implications in many human diseases, the understanding of DDX3X catalytic cycle remains limited. This understanding is crucial for a complete understanding of DDX3X role in cancer and viral propagation and for designing and development of novel cancer and viral therapeutic agents that target DDX3X function.

In addition to the RecA-like catalytic core, which is common to all members of DEAD-box family of enzymes, DDX3X possess N- and C-terminal domains.[13-15] The DDX3X catalytic core consists of residues 182 to 544, the DDX3X N-terminal consists of residues 1 to 181, and the DDX3X C-terminal consists of residues 545 to 662. The three-dimensional structure of DDX3X catalytic core, in complex with AMP, or a non-hydrolyzable ATP analog, adenylyl imidodiphosphate (AMPPNP), is known from crystallographic studies.[13-15] On the other hand, the structures of the N- and C-terminal domains of DDX3X are not known and are predicted to be unstructured.[13] A DDX3X construct containing residues 168 to 582, hence parts of the N-terminal and C-terminal domains, showed no ATPase activity[13] (Fig. 1). On the other hand, a construct containing residues 132 to 582 possesses ATPase activity.[15] Structural and biochemical data demonstrated that residues 152 to 160 of the DDX3X N-terminal domain are necessary for the RNA dependent ATPase activity of DDX3X.[14,15] Lastly, a DDX3X construct containing residues 132 to 607, demonstrated 10-fold increase in helicase activity when compared to the construct containing residues 132 to 582, suggesting that residues 583 to 607 enhance the helicase activity of DDX3X.[14] Here we investigate the nucleic acid structure requirement for ATPase activity, and the multimeric state of the DDX3X construct containing residues 1 to 582. Our construct is 80 residues shorter than the wild-type DDX3X (Fig. 1A). We were unable to purify the wild-type DDX3X protein in *Escherichia coli* (*E. coli*), at the large quantities necessary for enzymatic investigation. Since previous data suggest that residues 583 to 607 of the C-terminal enhance the DDX3X helicase activity,[14] it is possible our construct has reduced helicase activity when compared to wild-type DDX3X. Nevertheless, this construct is a good model system for the investigation of the RNA structural requirement and the DDX3X protein multimeric states needed to support the DDX3X ATPase activity. This information is important for the elucidation of the DDX3X protein's unique features and structural properties, which could be potential specific inhibitor targets.

**RESULTS**

The Enemark laboratory's experiments with a DDX3X construct containing residues 132 to 582, showed that single-stranded RNA and double-stranded RNA molecules do not support the ATPase activity of the 131 to 582 DDX3X construct.[15] Only molecules that contained a single-stranded–double-stranded RNA region supported the helicase activity of Enemark's group's DDX3 construct.[15] Since for a number of DEAD-box proteins, a single-stranded RNA is sufficient to support the ATPase activity of the protein,[1,16] we decided to investigate this observation further.

Figure 2 shows the dependence of the ATPase activity of DDX3X in presence of different nucleic acids substrates (Table 1). The ATPase activity of DDX3X, at various nucleic acid concentrations, was measured via thin-layer chromatography (TLC). RNA substrates A and B are the same RNA substrate as those used by the Enemark group.[15] The data in Fig. 2 show substrate B, which contains single-stranded–double-stranded RNA junctions, supports the ATPase activity of DDX3X. The RNA substrate consisting of only a double-helix, substrate A, did not support the helicase activity of DDX3X. None of the single-stranded RNA substrates D, E, F, H (Table 1) stimulated the ATPase activity of DDX3X. Hence, our data agree with the data from the Enemark group[15] suggesting that a single-stranded–double-stranded RNA junction is required to support the DDX3X catalytic activity.

DNA supports the catalytic activity of a small number of DEAD-box proteins.[17,18] We were interested in investigating if DDX3X's ATPase activity is stimulated by DNA. Thus, we studied the ability of a DNA substrate containing single-stranded–double-stranded regions (Table 1, substrate C) to stimulate DDX3X ATPase activity. DNA substrate C has the same sequence as RNA substrate B, which stimulates the ATPase activity of DDX3X. As data in Fig. 2 show, DNA substrate C does not stimulate the ATPase activity of DDX3X.

Interestingly, the data in Fig. 2 show that as the concentration of RNA substrate B increases, the extent of ATP hydrolyzed decreases. This suggests that DDX3X during its catalytic cycle may act as a multimer. When the concentration of RNA becomes large, the population of RNA molecules that have more than one DDX3X molecule bound to them decreases, and consequently the extent of ATPase activity of DDX3X decreases.

To test the above hypothesis and to decipher why certain RNA molecules supported the ATPase activity of DDX3X while others did not, we performed electrophoretic mobility shift assays (EMSA) in the presence of different RNA substrates. Figure 3A shows the dependence of fraction RNA shifted versus the DDX3X protein concentration for substrates A, B, D and G. Substrate B, which supports the ATPase activity of DDX3X, also supports its binding. Using the Hill equation to fit the binding data of DDX3X to substrate B we obtain a Hill coefficient of $2.47 \pm 0.06$. Hence, the binding of DDX3X to this substrate is cooperative. A Hill coefficient of larger than two but smaller than three means that three molecules of DDX3X are bound cooperatively to RNA substrate B.[19]

While a ladder of bands—due to the slower mobility within the gel of protein-nucleic acid complexes with increasing of numbers protein subunits—can sometimes be observed under favorable conditions, in a large number of experiments the multimeric and monomeric forms of protein interacting with nucleic acid cannot be resolved by EMSA[20]. In our experiments, we cannot discriminate between one monomer or multiple monomers of DDX3 bound to RNA; all different protein-RNA complexes migrate the same distance in the polyacrylamide gel.

Interestingly, single-stranded RNA substrate D, which does not stimulate the ATPase activity of DDX3X, supports DDX3X binding. However, this binding is less cooperative than the binding of DDX3X to substrate B. The Hill coefficient for binding of DDX3X to substrate D is $1.30 \pm 0.10$. Thus, only two molecules of DDX3X bind cooperatively to a single-stranded RNA. Combined the ATPase and the binding data obtained in presence of substrates B and D suggest DDX3X cooperative dimer assembly is not sufficient to stimulate the DDX3X ATPase activity, and assembly of a cooperative trimer is required to support this activity.

The maximum fraction of shifted RNA-DDX3X complex is smaller in presence of substrate D instead of substrate B (Table 1, Fig. 3A). There are two possible explanations for these results. First, a fraction of substrate D is refractory to DDX3X binding. We investigated the structure of substrate D using RNA folding web application Mfold[21]. As predicted from Mfold, construct D does not form stable secondary structures. Therefore, all the D molecules should support DDX3X binding to RNA in the same way. Another possibility is that substrate D in complex with DDX3X is less stable than the substrate C in complex with DDX3X, and a fraction of substrate D in complex with DDX3X is coming apart during electrophoresis[20,22].

RNA substrate A (Table 1), which contains only a double-helix, does not support DDX3X binding as investigated by EMSA. Therefore, this substrate is unable to stimulate ATP hydrolysis, because it does not support DDX3X binding. We investigated the ability of our DDX3X construct to bind to a longer double-stranded RNA substrate (Table 1, substrate G). The data in Fig. 3 show that DDX3X does not bind to this RNA substrate either. Thus, the DDX3X construct used in this study does not bind to a double-stranded only RNA structures.

Subsequently, we investigated the effect of the non-hydrolyzable ATP analog, AMPPNP, on the equilibrium binding of DDX3X to different RNA substrates. Figure 3B shows the fraction of RNA bound to DDX3X versus DDX3X concentrations in presence of 6 mM AMPPNP. The double-stranded RNA substrates A and G, which did not support DDX3X binding in absence of AMPPNP, did not support the DDX3X binding when AMPPNP was present either. Substrate D supported the DDX3X binding with the same affinity and cooperatively in presence or absence of AMPPNP. Interestingly, substrate B supported the binding of DDX3X with a Hill coefficient of $2.06 \pm 0.08$ in presence of AMPPNP, instead of $2.47 \pm 0.06$ when AMPPNP was absent. Therefore, the binding of AMPPNP to two RNA associated monomers, prevents the cooperative binding of third monomer. Alternatively, three DDX3X monomers bind to RNA, upon subsequent AMPPNP binding to the trimer, one of the monomers dissociates.

We investigated the interaction of DDX3X monomers with RNA by using chemical cross-linking and gel electrophoresis. DDX3X monomers were cross-linked to RNA and each other using 1-ethyl-3-(3 dimethylaminopropyl)carbodiimide (EDC). EDC activates the carboxylic groups in proteins for interaction with proteins' or nucleic acids' primary amines. The end product of the reaction is the formation of the amide bonds between the carboxylic groups and primary amines.[23]

Substrates B and D were utilized for the EDC cross-linking studies. To visualize the protein-RNA complex, the RNA molecules were $^{32}$P-labeled. As suggested from the EMSA data, DDX3X binds to both substrates B and D; consequently, it also cross-links to those substrates (Fig. 4). Figure 4 shows that at least three molecules of DDX3X are cross-linked to each RNA substrate. The DDX3X monomer has a molecular mass of 67.7 kDa, the RNA substrate B has a molecular mass of 18.6 kDa, and the RNA substrate D has a molecular mass of 9.4 kDa. Based on the molecular mass of DDX3X monomer and RNA molecules, the DDX3X-RNA cross-linked monomers and dimers migrate at the correct molecular mass on the denaturing gel, instead the trimer migrates slower that its molecular mass. The molecular mass of the trimer cross-linked to either RNA substrate B and D should be 221.7 and 212.5 kDa respectively, instead these trimers migrate as molecules with a mass of 270 kDa. The discrepancy of trimers' migrations could be a consequence of the fact that the gel we used to perform these experiments was a 3–15% polyacrylamide gradient gel, and the resolution of the gel at low polyacrylamide concentration, where the DDX3X trimer resides, is not very good. It is also possible that the cross-linking of DDX3X monomers to each other and RNA makes it more difficult for the complex to serpent through the gel when compared with a non-cross linked globular protein of the same mass. In fact, the retardation of the UV protein-RNA complexes migration on SDS page gels has been previously observed.[24]

Hence, the cross-linking data suggests that both substrates B and D support the binding of at least three DDX3X monomers. Combined the cross-linking and the EMSA data suggest that difference in binding of DDX3X to substrate B and D is that three DDX3X monomers bind to substrate B in a cooperative manner, instead two DDX3X monomers bind to substrate D cooperatively. The observation that at least three molecules of DDX3X are cross-linked to the single-stranded substrate D also suggests that DDX3X monomers, in addition to interacting cooperatively with each other in the presence of an RNA substrate, could also load onto a single-stranded region uncooperatively.

The question remains what structural elements of substrate B support the cooperative interaction of three DDX3X monomers. It could not be the double-stranded region; the double-stranded RNA, substrate A (Table 1), which has the same sequence as that of the double-stranded region in substrate B, does not support DDX3X binding (Fig. 3A and 3B). It could not be the single-stranded region, the single-stranded substrate D, with very similar sequence to substrate B, does not support cooperative binding of three DDX3X monomers (Fig. 3A and Table 2). Therefore, it must be the single-stranded–double-stranded regions in substrate B that supports the assembly of DDX3X trimer. Since only substrates containing a single-stranded–double-stranded region support the ATPase activity of DDX3X containing residues 132 to 582[15] and the DDX3X construct used in this study, then the DDX3X cooperative trimer assembly supported by the single-stranded–double-stranded RNA region is required for the DDX3X ATPase activity.

Lastly, we investigated if the stimulation of ATP hydrolysis by substrate B was a cooperative process. Thus, we investigated, via TLC, the fraction of ATP hydrolyzed versus DDX3X concentration. As the data in Fig. 5 show, the dependence of the ATP hydrolyzed on the protein concentration has a sigmoid shape, suggesting allosteric interactions. Hence, the Hill equation, instead of the Michealis-Menten equation was used to fit the data. The Hill constant obtained from these data is $1.99 \pm 0.19$ (Table 2). Therefore, while three monomers of DDX3X are implicated in the ATPase cycle, only two molecules of ATP are cooperatively hydrolyzed during this process.

**DISCUSSION**

In this work, we investigate the action mechanism of a DDX3X construct that is missing the last 80 residues of its C-terminal domain. Different from many DEAD-box RNA helicases for which a single-stranded RNA region or a blunt-ended RNA double-helix is sufficient to stimulate their ATPase activity,[1,16,25] only an RNA molecule that contains a single-stranded–double-stranded region supports the ATPase activity of DDX3X. This conclusion agrees with the previous experiments by the Enemark group performed with a shorter DDX3X construct, which lacked 80 residues from C-terminal domain and 132 residues from the N-terminal domain.[15] In addition, our data suggests that the RNA substrate containing the single-stranded–double-stranded region supports DDX3X cooperative trimer assembly, the single-stranded RNA substrate only supports assembly of a cooperative dimer, and a blunt-ended RNA double-helix does not support DDX3X binding. Thus, the formation of the DDX3X trimer is supported by the single-stranded–double-stranded RNA region and is required for the DDX3X ATPase activity.

Previous single-molecule fluorescence experiments have suggested that LAF-1, DDX3X ortholog in *Caenorhabditis elegans* (*C. elegans*) binds to a single-stranded–double-stranded RNA molecule as a multimer.[26] In addition, experiments performed with Ded1, the DDX3X ortholog in *Saccharomyces cerevisiae* (*S. cerevisiae*)**,** have suggested that Ded1 forms a trimer when alone and in presence of an RNA molecule having a single-stranded–double-stranded region. Thus, our conclusions concerning the multimeric state of DDX3X and the RNA structure required to form this state agree with the previous data on the DDX3X ortholog in *C. elegans* and *S. cerevisiae.*

Interestingly, the dependence of ATP hydrolysis on protein concentration suggests that only two molecules of ATP are hydrolyzed cooperatively by the three DDX3X monomers. There are four possible logical reasons why only two molecule of ATP are hydrolyzed cooperatively by three DDX3X monomers: (i) only two ATP molecules are bound to the DDX3X trimer; hence, one of the DDX3X monomers has an unoccupied ATP pocket, (ii) three ATP molecules are bound to the DDX3X trimer, but only two of them are hydrolyzed, the third ATP is never hydrolyzed, (iii) three ATP molecules are bound to the DDX3X trimer but only two are hydrolyzed cooperatively, (iv) the catalytically active unit of the DDX3X multimer is a dimer in complex with two ATPs. The trimer is initially assembled, but upon ATP binding, one of the monomers dissociate from the assembled complex.

The equilibrium binding data of DDX3X to substrate B in the presence of AMPPNP imply that only two monomers of DDX3X bind cooperatively under these conditions (Table 2). This observation suggests two possible modes of interaction of DDX3X molecules with each other and their substrate. First, the binding of two monomers in complex with non-hydrolyzable ATP to RNA could prevent the formation of the stable cooperative trimer. In other words, a DDX3X trimer could not assemble if both the first and the second RNA binding monomers of DDX3X contain bound ATP. Second, a cooperative trimer could assemble initially, but once two or more ATP molecules bind to the trimer, one of the monomers dissociate from the complex. The trimer assembly could be required for the initial loading of DDX3X to the RNA single-stranded–double-stranded region, but not for its catalytic activity. If the first mode of interaction is correct, then the observation that two ATP molecules hydrolyze cooperatively during the DDX3X catalytic cycle could be a consequence of two molecules of ATP being bound to the DDX3X trimer, where the middle bound monomer has no ATP bound. If the second mode of interaction is correct, the observation that two molecules of DDX3X are hydrolyzed cooperatively during the DDX3X catalytic cycle could be a consequence of a DDX3X dimer with two bound ATP molecules being the DDX3X functional unit. The data presented here cannot discriminate between these two models and experiments are underway to decipher the intricate kinetics of DDX3X, RNA and ATP interactions. Nevertheless, our data suggest that a cooperative DDX3X trimer in complex with three ATP molecules is not the DDX3X active catalytic complex.

Similar to DDX3X, the three monomers of Ded1 hydrolyze cooperatively two molecules of ATP during the catalytic cycle.[27] For Ded1, it was suggested that only two of the three monomers of Ded1 possess ATP during the catalytic cycle.[27]

The dissociation constant ($K_d$) of DDX3X binding to RNA substrate B as measured by the TLC ATPase assay is $551 \pm 50$ nM, while the $K_d$ as measured by EMSA is $285 \pm 18$ nM (Table 2). These differences could imply that the DDX3X binding constant as measured from the ATPase assay contains contributions from other kinetic steps. Differences between the RNA binding affinity as measured via a direct assay such as EMSA and measured indirectly by an ATPase assay have been observed for other members of DEAD-box family of enzymes.[28] Alternatively, the differences between the $K_d$ values could be a consequence of the fact that EMSA assay detects all the DDX3X–RNA complexes—monomer, dimer and trimer—while the ATPase assay detects only assembly of the catalytic active complex, which our data suggest could be a trimer or a dimer. The $K_d$ of multimer binding to RNA would be larger than that of the monomer binding.

EMSA experiments show that the blunt-ended substrate A (Table 1) did not support the binding of the DDX3X construct used in this study (Fig. 3A). Our DDX3X construct possessed residues 1 to 582. However, substrate A supported the binding of the DDX3X construct used in a previous study, which lacked most of the N-terminal residues.[15] Combined, our results and the previous data suggests that the residues in the N-terminal of DDX3X may have a role in deciding the DDX3X RNA binding region. In fact, the wild-type LAF-1 construct possessing its N-terminal domain, as measured by EMSA assay, does not bind to an RNA double-helix.[26] Hence, it is possible that DDX3X and LAF-1 N-terminal domain roles are to prevent these proteins from interacting with double-stranded RNA regions. Many DEAD-box proteins use ancillary domains to guide them to their site of action and dictate the double-helix substrates those proteins act upon.[29,30]

While the ATPase and the helicase activity of very few DEAD-box proteins is activated by DNA,[17,18] the ATPase activity of the DDX3X construct used in this study is not activated by the DNA molecule we used. The DNA molecule contained single-stranded–double-stranded region and had the same sequence as the RNA molecule that activated the ATPase activity of DDX3X. In addition, the ATPase activity of Ded1 is not activated either by single-stranded or double-stranded DNA molecules,[31] and the helicase activity of Ded1 is not supported by DNA substrates containing single-stranded–double-stranded junctions.[32] Interestingly, the helicase activity of Ded1 is supported by DNA–RNA chimera constructs and hybrid RNA/DNA molecules[32]. These constructs were not investigated in this study; hence, it is possible that the catalytic activity of DDX3 may be activated by DNA–RNA chimeras. Nevertheless, combined previous data and the data shown here suggest that the inability to act on a DNA-only molecule could be a property of DDX3X family of enzymes.

**CONCLUSIONS**

Similar to most of the members of DEAD-box family of enzymes, DNA does not support DDX3X ATPase activity.[1,16] On the other hand, DDX3X is different from most DEAD-box family members in its protein and RNA requirement for catalytic activity. DDX3X, like Ded1, its ortholog in *S. cerevisiae*, forms a cooperative trimer in presence of an RNA molecule containing single-stranded–double-stranded region, and this trimer is required for DDX3X catalytic activity.[27] To our knowledge, Ded1 and DDX3X are the only RNA helicases for which a trimer formation is required for catalytic activity. Thus, the protein surface between the monomers is an ideal target for specific DDX3X inhibitors target. These inhibitors would halt DDX3X catalytic cycle by preventing trimer assembly. Once the *in vivo* RNA substrates of DDX3X are identified, the single-stranded–double-stranded RNA regions on those substrates could also server as specific DDX3X inhibitors.

**MATERIALS AND METHODS**

**Materials**

All chemicals were bought from Thermo Fisher Scientific. RNA substrates were commercially obtained and HPLC purified from Integrated DNA Technologies. γ-$^{32}$P-labeled ATP was obtained from PerkinElmer. PEI Cellulose F coated TLC plates were bought from EMD Millipore.

**Protein expression and purification**

The pNIC28 vector bearing the complete sequence of the wild-type DDX3X protein and an N-terminal His-tag with a TEV cleavage site (Fig. 1, Supplementary Fig. S1) was a gift from the Karolinska Institute. The wild-type DDX3X contains 662 amino acids and consists of the N-terminal domain, catalytic core and C-terminal domain (Fig. 1A, Supplementary Fig. S1). We substituted the sequence of amino acid 583 in the DDX3X coding sequence with that of a stop codon. Hence, our DDX3X construct is missing 80 C-terminal residues (Fig. 1A). Protein expression and purification was carried out as specified by Högbom, et al.[13] In brief, the DDX3X construct bearing an N-terminal His-tag was expressed in *E. coli* C2566I (NEB). Nickel affinity column (HisPur Ni-NTA Superflow Agarose, Thermo Scientific) and size-exclusion column (Sephacryl S-200HR, GE Healthcare Lifesciences) were used to purify the protein. The His-tag was not removed from DDX3X.

**TLC ATPase assay**

Hydrolysis of ATP was monitored using TLC as previously described[33]; however, γ-$^{32}$P-labeled ATP was used instead of α-$^{32}$P-labeled ATP. Two kinds of TLC ATPase assay were performed, one where the protein concentration was kept constant and the RNA concentration was varied, and the other one in which the RNA concentration was kept constant and the protein concentration was changed. For the assays in which the RNA concentration was varied, first the RNA was annealed by incubating it with 50 mM HEPES-KOH pH 7.5, 50 mM KCl at 95°C for 1 min, 65°C for 3 min, cooling it down to 25°C for 1 min, and adding 10 mM MgCl$_2$ final. Subsequently, different RNA concentrations were mixed with 50 mM HEPES-KOH pH 7.5, 1.3 mM MgCl$_2$, 50 mM KCl, 2 mM DTT, 1 μM protein, 0.5 mM ATP, 2.5 nM γ-$^{32}$P-labeled ATP and incubated at 37°C for 30 min. After incubation, the reactions were quenched with 0.5 M EDTA pH 8, spotted on PEI Cellulose F coated TLC plates. The TLC plates were developed using a solvent system consisting of 0.75 M LiCl and 1 M acetic acid. TLC plates were left to dry, then exposed to phosphor imaging screens. The screens were scanned using a Personal Molecular Imager System (Bio-Rad Laboratories), and analyzed using Quantity One (Bio-Rad Laboratories). The fraction of ATP hydrolyzed was calculated as the ratio of the counts on the inorganic phosphate band over the total counts on the lane.

Assays where the protein concentration is varied were performed similar to the assays where RNA was varied, with the following changes: 2 mM ATP, 0.01 µM γ-$^{32}$P-labeled ATP, 1 nM RNA were present in the reaction mixture of these assays. The fraction of ATP hydrolyzed versus protein concentrations were fit to the Hill equation (1) using OriginLab:

$$f_B = f_B(0) + [f_B(\max) - f_B(0)]\left\{\frac{[protein]^n}{K_d^n + [protein]^n}\right\} \quad (1)$$

where $f_B$ is the maximum amount of ATP hydrolyzed, $f_B(0)$ and $f_B(\max)$ are the lower and upper baselines of the binding curve, $K_d$ is the dissociation constant, and $n$ is the Hill coefficient.

**RNA binding gel-shift assay**

RNA substrates were annealed as described for the TLC ATPase assay. 1 nM final annealed RNA was incubated with different concentrations of DDX3X in 50 mM HEPES-KOH pH 7.5, 50 mM KCl, 1.3 mM MgCl$_2$, 2 mM DTT, 20% glycerol in presence or absence of 6 mM final AMPPNP·Mg. The reaction mixture was incubated for 30 minutes at 22°C and then loaded on a non-denaturing gel. The non-denaturing gel was 10% polyacrylamide with a ratio of acrylamide to bis-acrylamide of 29:1. The gel buffer and the running buffer consisted of 0.33X TBE buffer and 5 mM MgCl$_2$. The gels were run at 22°C for 2 hours at a voltage of 10 V·cm$^{-1}$. Next, the gels were dried in a Gel Dryer (Model 583, Bio-Rad Laboratories), and exposed to a phosphor-imaging screen. Exposures were imaged using a Personal

Molecular Imager System (Bio-Rad Laboratories), and analyzed using Quantity One (Bio-Rad Laboratories). The Hill equation (2) was used to fit the data:

$$f_B = f_B(0) + [f_B(\max) - f_B(0)]\left\{\frac{[protein]^n}{K_d^n + [protein]^n}\right\} \quad (2)$$

In equation 2, $f_B$ is the fraction of RNA bound to protein, $f_B(0)$ and $f_B(max)$ are the lower and upper baselines of the binding curve, $K_d$ is the dissociation constant, and $n$ is the Hill coefficient. The program used to fit the data was Originlab.

**Protein-RNA crosslinking**

Cross-linking reactions were carried out similarly to protein-RNA binding reactions. Reactions, contained 1 μM protein, and 10 nM of $^{32}$P-labeled RNA substrate. The RNA substrates were annealed using similar conditions as described for the malachite green ATPase assays. Reactions constituents were combined on ice, then incubated for 30 minutes at room temperature to allow reactions to equilibrate. Following room-temperature equilibration, 2 μL of 200 mM EDC (20 mM final) was added to the reactions, and incubated at room temperature for 45 minutes. Reactions were quenched by adding 5X SDS-PAGE loading buffer. Next, the mixture was boiled at 95°C for 2 minutes, and loaded onto a 3–15% polyacrylamide Tris-acetate gel, prepared and ran as previously described by Cubillos-Rojas, et al.[34]

## ACKNOWLEDGEMENTS

The authors thank Apostolos Gittis for many stimulating discussions. This work was supported by NIH [grant number 5R21CA175625-02 to E. K.] and start-up funding from the Chemistry Department at the University of Central Florida.

## AUTHOR CONTRIBUTION STATEMENT

Anthony F. T. Moore, Aliana López de Victoria and Eda Koculi designed experiments. Anthony F. T. Moore and Aliana López de Victoria performed the experiments. Anthony F. T. Moore and Eda Koculi wrote the paper.

**COMPETING FINANACIAL INTERESTS**

The authors declare no financial conflicts of interest.

# TABLES

**TABLE 1.** Sequences of RNA substrates used in this study.

| RNA Substrates | Sequences |
|---|---|
| A | ```
5'-GGCGGCCGCC-3'
   ||||||||||
3'-CCGCCGGCGG-5'
``` |
| B | ```
5'-(U)₂₀-GGCGGCCGCC-3'
         ||||||||||
      3'-CCGCCGGCGG-(U)₂₀-5'
``` |
| C | ```
5'-(dT)₂₀-dGdGdCdGdGdCdCdGdCdC-3'
          |  |  |  |  |  |  |  |  |  |
       3'-dCdCdGdCdCdGdGdCdGdG-(dT)₂₀-5'
``` |
| D | 5'-(U)₂₀-GGUGGUUGUG-3' |
| E | 5'-CUCAACUCAA-3' |
| F | 5'-CUCAACUCAACCCUUCAU-3' |
| G | ```
5'-CUCAACUCAACCCUUCAUCUCAACUCAACCCUUCAU-3'
   ||||||||||||||||||||||||||||||||||||
3'-GAGUUGAGUUGGGAAGUAGAGUUGAGUUGGGAAGUA-5'
``` |
| H | 5'-(C)₁₈-3' |

**TABLE 2.** Equilibrium parameters of DDX3X–substrate interactions.

| RNA substrate | Parameter | DDX3X–RNA binding | |
|---|---|---|---|
| | | Nucleotide | |
| | | −AMPPNP | +AMPPNP |
| A | $n$ | - | - |
| | $K_d$ (DDX3X, nM) | ~0 | ~0 |
| B | $n$ | 2.47 ± 0.06 | 2.06 ± 0.08 |
| | $K_d$ (DDX3X, nM) | 285 ± 18 | 114 ± 10 |
| D | $n$ | 1.30 ± 0.10 | 1.17 ± 0.01 |
| | $K_d$ (DDX3X, nM) | 298 ± 31 | 217 ± 60 |
| G | $n$ | - | - |
| | $K_d$ (DDX3X, nM) | ~0 | ~0 |

[1]$K_d$ was determined by fitting the Hill equation to the EMSA data. The values shown are the averages values obtain by a minimum of two independent experiments, and the errors are the standard deviations from those averages.

1.

**FIGURE LEGENDS**

**FIGURE 1.** DDX3X linear diagram**.** The construct used in this study contains a terminal His-tag and a TEV cleavage site. The His-tag and the TEV cleavage site region is shown as a dotted area. The DDX3X N-terminal in shown in cyan, the catalytic core in red, and the C-terminal in green. The numbering starts with the first amino acid of DDX3X protein sequence. The DDX3X construct used in this study is 80 amino acids shorter than wild-type DDX3X.

**FIGURE 2.** Stimulation of DDX3X ATPase activity by various RNA and DNA substrates. The plot shows fraction of ATP hydrolyzed versus nucleic acid concentration. The structure and the sequence of the RNA and DNA substrates employed are shown in Table 1. The average data points are connected by linear segments. The error bars are the standard deviation from at least two independent experiments. Legend: RNA substrate A (—●—); RNA substrate B (—■—); DNA substrate C (—□—), RNA substrate D (—◆—); RNA substrate E (—◇—); RNA substrate F (—△—); RNA substrate H (—○—). RNA substrates' sequence and structure are shown in Table 1.

**FIGURE 3.** Equilibrium binding of DDX3X to various RNA substrates. A) Representative plots of DDX3X binding to different RNA substrates in absence of nucleotide; B) Representative plots of DDX3X binding to RNA in presence of AMPPNP. Legend: substrate A (—●—); substrate B (—■—); substrate D (—◆—); substrate G (—▼—). Substrates' sequence and structure shown in Table 1. The average values for $K_d$ and standard deviations are shown in Table 2.

**FIGURE 4.** EDC crosslinking of DDX3X in presence of either RNA substrate B or D (Table 1). This is a Tris-acetate gradient gel of 3–15% polyacrylamide. No cross-linking is observed in absence of protein (lanes 1 and 3). A stained protein ladder was used as a marker. The experiment was repeated four times.

**FIGURE 5.** Dependence of the ATPase activity of DDX3X on protein concentration. Substrate used for this experiment is B (—■—), which contains single-stranded–double-stranded-regions. The plot is representative of a single experiment. Hill equation was used to fit the data. The Hill coefficient obtained from the data fit was 1.99 ± 0.19, and the dissociation constant 551 ± 50 nM. The average values and the standard deviation are from two independent experiments.

**FIGURES**

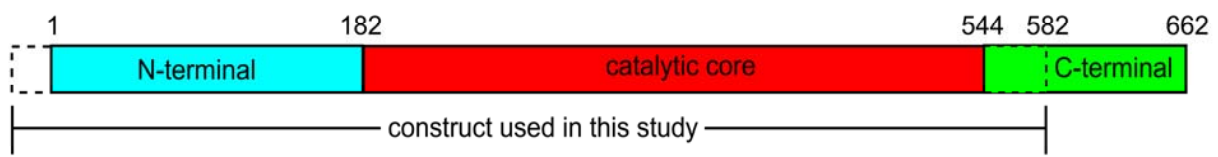

FIGURE 1

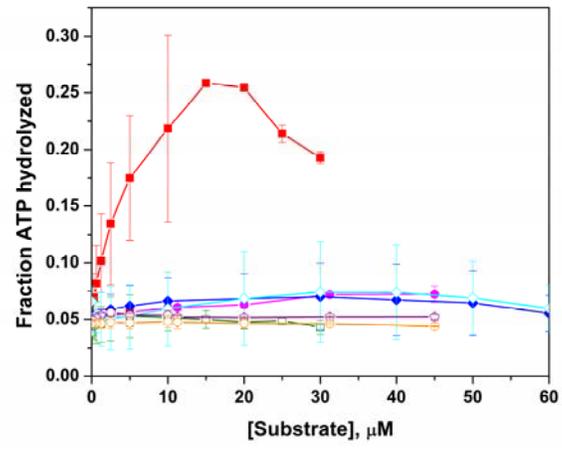

**FIGURE 2**

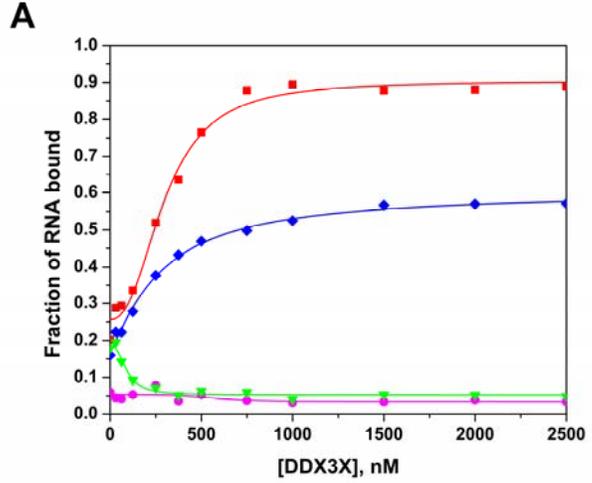

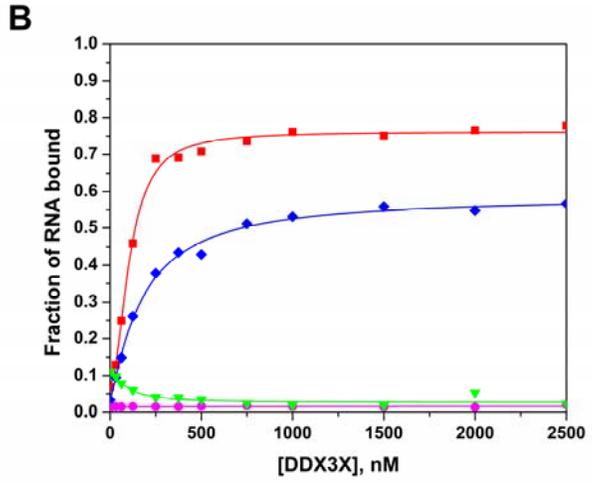

**FIGURE 3**

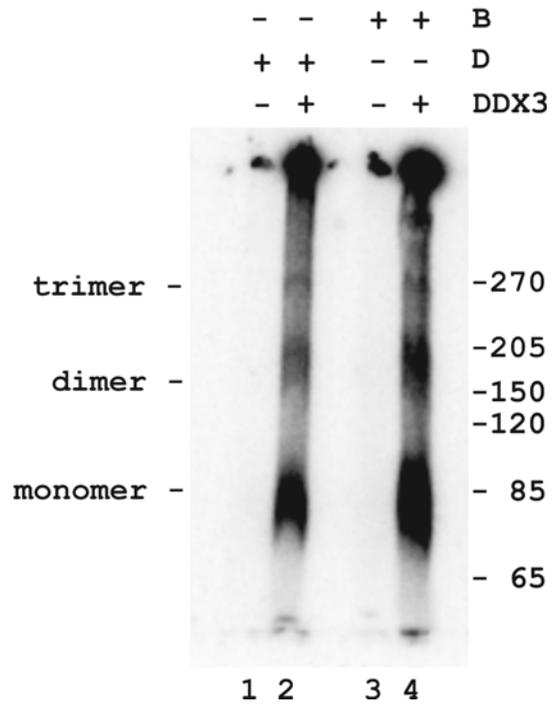

**FIGURE 4**

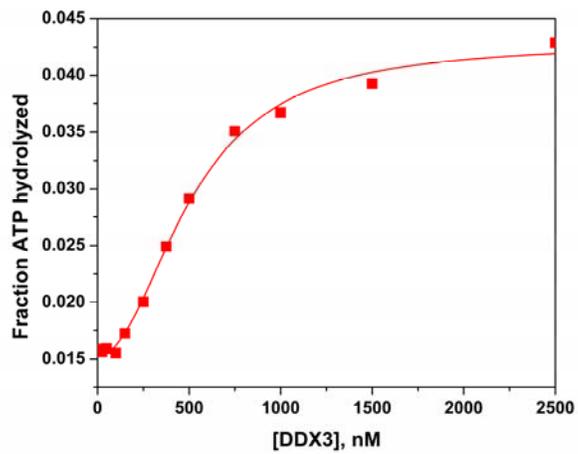

**FIGURE 5**

**SUPPLEMENTARY INFORMATION FOR:**

# Action mechanism of DDX3X: An RNA helicase implicated in cancer propagation and viral infection

Anthony F. T. Moore, Aliana López de Victoria, Eda Koculi[a]

University of Central Florida, Department of Chemistry, 4111 Libra Dr., Physical Sciences Bld. Rm. 255, Orlando, FL 32816-2366

[a.] Corresponding author. E-mail: eda.koculi@ucf.edu; Phone: (407)823-5451.

SUPPLEMENTAL MATERIALS FIGURE CAPTIONS

**SUPPLEMENTARY FIGURE S1.** The amino acid sequence of the DDX3X construct used in this study. Selected features, as highlighted in the amino acid sequence: N-terminal domain (cyan); catalytic core (red); residues 545-582 of the C-terminal domain (green). The non-colored region is the His-tag and the TEV cleavage sites.

**SUPPLEMENTARY FIGURE S2.** EMSA gels. Concentration of DDX3X construct increases from left-to-right. (A) RNA substrate A, in the absence of nucleotide; (B) RNA substrate A, in the presence of AMPPNP; (C) RNA substrate B, in the absence of nucleotide; (D) RNA substrate B, in the presence of AMPPNP; (E) RNA substrate D, in the absence of nucleotide; (F) RNA substrate D, in the presence of AMPPNP; (G) RNA substrate G, in the absence of nucleotide; (H) RNA substrate G, in the presence of AMPPNP.

```
MHHHHHHSSGVDLGTENLYFQSMSHVAVENALGLDQQFAGLDLNSSDNQSGGSTASKGRYIPPHLRNREA
TKGFYDKDSSGWSSSKDKDAYSSFGSRSDSRGKSSFFSDRGSGSRGRFDDRGRSDYDGIGSRGDRSGFGK
FERGGNSRWCDKSDEDDWSKPLPPSERLEQELFSGGNTGINFEKYDDIPVEATGNNCPPHIESFSDVEMG
EIIMGNIELTRYTRPTPVQKHAIPIIKEKRDLMACAQTGSGKTAAFLLPILSQIYSDGPGEALRAMKENG
RYGRRKQYPISLVLAPTRELAVQIYEEARKFSYRSRVRPCVVYGGADIGQQIRDLERGCHLLVATPGRLV
DMMERGKIGLDFCKYLVLDEADRMLDMGFEPQIRRIVEQDTMPPKGVRHTMMFSATFPKEIQMLARDFLD
EYIFLAVGRVGSTSENITQKVVWVEESDKRSFLLDLLNATGKDSLTLVFVETKKGADSLEDFLYHEGYAC
TSIHGDRSQRDREEALHQFRSGKSPILVATAVAARGLDISNVKHVINFDLPSDIEEYVHRIGRTGRVGNL
GLATSFFNERNINITKDLLDLLVEAKQEVPSWLENMAYEHHYKG
```

**FIGURE S1**

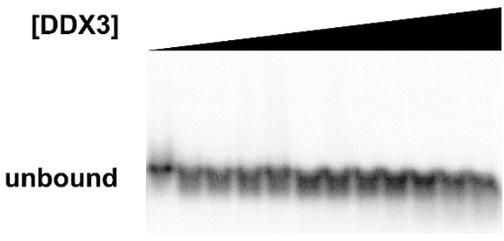
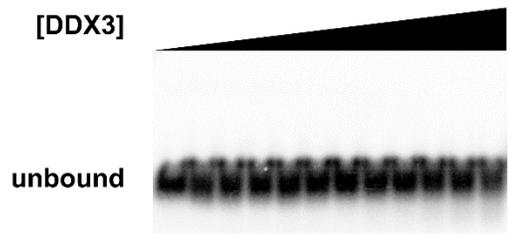
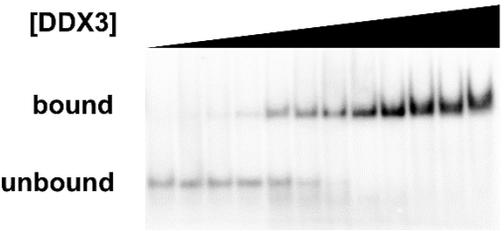
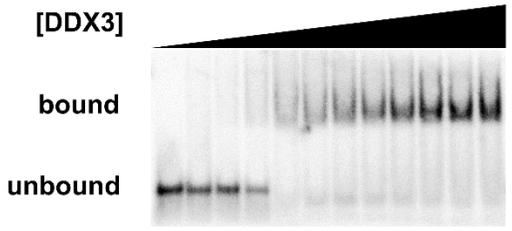
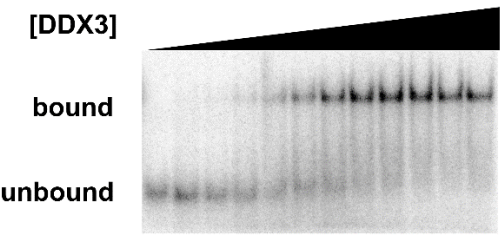
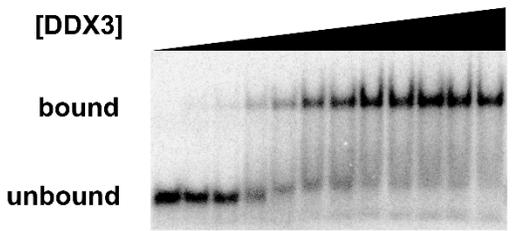
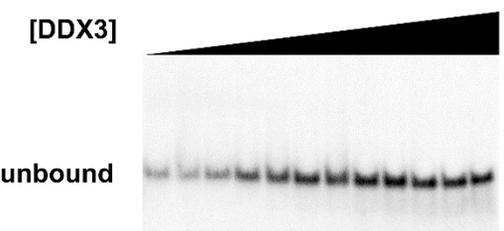
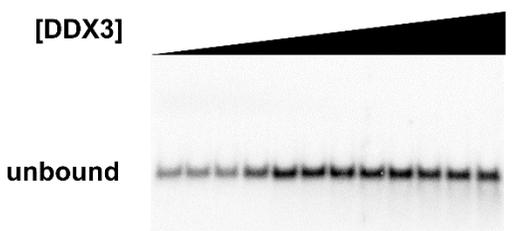

**FIGURE S2**